\documentclass[%
 preprint,
 amsmath,amssymb,
 prl,
]{revtex4-2}

\usepackage{graphicx}
\usepackage{dcolumn}
\usepackage{bm}
\usepackage{hyperref}
\usepackage{float}

\begin{document}

\preprint{APS/123-QED}

\title{Strain-transport superposition in shear-thinning dense non-Brownian suspensions}

\author{Rishabh V. More}
 \affiliation{Department of Chemical and Biological Engineering, Monash University, Clayton, Australia}
 \email{rishabh.more@monash.edu}




\date{\today}

\begin{abstract}
Shear thinning in dense non-Brownian suspensions is commonly attributed to shear-induced microstructural evolution, including changes in structural alignment, anisotropy, and near-contact statistics, yet how these structural changes influence particle-scale dynamics remains unclear. Using particle-resolved simulations of dense suspensions that shear thin through diverse microscopic mechanisms—including short-range attraction, repulsion, and load-dependent friction—we show that the magnitude of non-affine particle velocities is controlled solely by the imposed shear rate, independent of coordination number, structural anisotropy, and interaction details. In contrast, macroscopic stress and viscosity remain strongly sensitive to the underlying interaction mechanism. When mean-squared displacements measured transverse to the flow are rescaled by accumulated strain and by the non-affine velocity variance, data from all cases collapse onto a single master curve, revealing strain-controlled transport with a robust crossover from ballistic to diffusive dynamics. These results demonstrate a fundamental decoupling between particle-scale kinematics and macroscopic rheology in shear-thinning dense suspensions and identify non-affine velocity fluctuations as the emergent dynamical scale governing shear-driven particle transport.
\vspace{-6mm}
\end{abstract}

\maketitle
\vspace{-6mm}

\paragraph{Introduction.} Shear thinning—the decrease of viscosity $\eta$ with increasing shear rate $\dot{\gamma}$—is a ubiquitous feature of driven soft matter, spanning polymeric liquids, colloidal dispersions, emulsions, and dense suspensions \cite{ewoldt2022designing}. In dense \emph{non-Brownian} suspensions, shear thinning is commonly observed at high (but well below jamming) volume fractions and often precedes shear thickening or shear jamming at higher stress or under different particle-scale constraints \cite{Ellero2022ShearThinning, chatte2018shear}. While shear thickening now admits a quantitative microscopic framework based on stress-activated constraints and frictional contact networks \cite{GuazzelliPouliquen2018JFM,Morris2020ARFM,NessAndreottiSun2022ARCMP,WyartCates2014PRL,SetoMariMorrisDenn2013PRL}, the corresponding organizing principle for shear thinning remains less clearly established \cite{Ellero2022ShearThinning}.

Classically, shear thinning has been attributed to flow-induced microstructural reorganization, including changes in near-contact statistics, anisotropic pair correlations, and the formation or breakup of transient structures that alter dissipation pathways \cite{Hoffman1972TSR,BradyMorris1997JFM,WagnerBrady2009PT, vazquez2016shear}. However, dense suspensions with different microscopic interactions—attraction, repulsion, or friction—often exhibit comparable shear-thinning trends despite exhibiting markedly different steady-state coordination, anisotropy, and force networks \cite{GuazzelliPouliquen2018JFM, more2021unifying, khan2023thinning}. This raises a fundamental question: is shear thinning governed primarily by interaction-specific microstructural relaxation, or by some other physics?

At the particle scale, shear-driven suspensions display strong non-affine motion and correlated rearrangements, leading to shear-induced dispersion even in the absence of thermal noise \cite{LeightonAcrivos1987JFM,BradyMorris1997JFM}. In shear-thickening and near-jamming regimes, these kinematic fluctuations become tightly coupled to stress transmission through growing contact networks and correlated clusters \cite{SetoMariMorrisDenn2013PRL,SinghNessMorris2020PRL, singh2023scaling, ClavaudSingh2025RheolActa}. By contrast, in shear-thinning states, the particle network reorganizes without an obvious signature in particle-scale transport. It remains unclear whether non-affine motion in this regime is controlled by interaction-specific mechanisms or by a different, more general dynamics.

In this Letter, we establish a kinematic framework for shear thinning in dense non-Brownian suspensions. Using particle-resolved simulations under steady shear at a rate $\dot{\gamma}$, we systematically vary microscopic interaction mechanisms responsible for shear-thinning—including attractive \cite{singh2019yielding}, repulsive \cite{chatte2018shear, more2021unifying}, and load-dependent frictional interactions \cite{more2020effect, khan2023rheology, lobry2019shear}—and measure microstructure, velocity correlations, non-affine velocity fluctuations $\langle|\mathbf{v}_{\mathrm{na}}|^2\rangle$, and particle transport. Results presented identify non-affine velocity fluctuations as the emergent dynamical scale governing shear-driven particle transport, leading to a \emph{strain-transport superposition} that demonstrates a fundamental decoupling between kinematics and stress in shear-thinning dense non-Brownian suspensions. The simulation setup and microstructural metric probed are described in the Appendix, with full details provided in the Supplemental Information (SI), which together support and complete the results and discussion presented here.


\begin{figure*}[t]
\includegraphics[width=\textwidth]{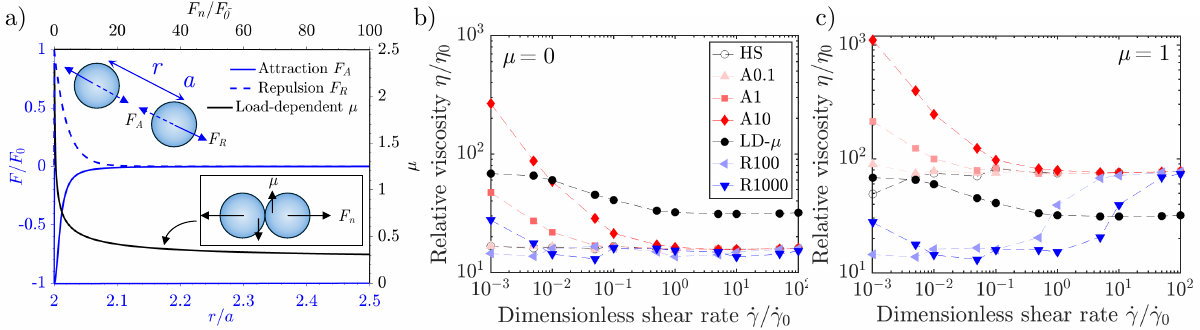}
\vspace{-3mm}
\caption{\label{fig:fig1}
(Color online) Microscopic interaction models and resulting shear-thinning rheology in dense suspensions. 
(a) Schematic of the short-range pairwise force laws used in the simulations, including attractive and repulsive interactions of varying strength and a load-dependent friction model. 
(b) Relative viscosity $\eta_r=\eta/\eta_0$ as a function of dimensionless shear rate $\dot{\gamma}/\dot{\gamma}_0$ for frictionless particles ($\mu=0$). 
(c) Relative viscosity $\eta_r$ versus $\dot{\gamma}/\dot{\gamma}_0$ for frictional particles ($\mu=1$). 
Despite their distinct microscopic origins, all interaction models exhibit pronounced shear thinning, while differing substantially in the magnitude and shear-rate dependence of the stress response. Lines in (b) and (c) are guides to the eye only.}
\vspace{-4mm}
\end{figure*}

\paragraph*{\textbf{Interaction models and force scales.}} All interaction strengths between particles of size $a$ are expressed relative to the characteristic hydrodynamic force scale $F_0 = \eta_0 \dot{\gamma}_0 a^2$, which also sets the characteristic shear rate $\dot{\gamma}_0$, where $\eta_0$ is the (constant) Newtonian solvent viscosity. Rate dependence arises from competition between $F_H \sim  \eta_0 \dot{\gamma} a^2$ and non-hydrodynamic forces, each introducing a stress scale $\sim |F_\alpha|/a^2$. Short-range repulsion (R) and attraction (A) act normal to particle surfaces with fixed range $\lambda=0.02a$ and magnitudes $F_R$ and $F_A$ as depicted in Fig.~\ref{fig:fig1}a; we study $F_A/F_0=\{0.1,1,10\}$ and $F_R/F_0=\{100,1000\}$ to vary strength of these interactions from weak, moderate to strong.
Contact forces are activated upon overlap and follow a Hertzian normal law with Coulomb friction \cite{luding2008cohesive}. Except for the normal load-dependent ($F_n$) friction case (LD-$\mu$), the friction coefficient is constant with $\mu=0$ or $\mu=1$; in LD-$\mu$,
\begin{equation}
\mu(F_n)=0.27\,\coth\!\left[0.27\left(\frac{F_n}{F_0}\right)^{0.35}\right],
\end{equation}
so that friction weakens with increasing normal load \cite{more2020effect, khan2023rheology, lobry2019shear}. All interaction cases are summarized in Table~\ref{tab:interaction_cases} with mathematical details in SI. Fig.~\ref{fig:fig1}(a) illustrates the short-range pairwise force laws used to model attractive and repulsive interactions of varying strength, as well as a load-dependent friction model. Although these interaction rules correspond to distinct physical mechanisms, all give rise to shear thinning in dense suspensions.

\begin{table}[b]
\vspace{-6mm}
\centering
\caption{Interaction cases. Forces are scaled by $F_0=\eta_0\dot{\gamma}a^2$. The repulsive interaction has Debye length $\lambda=0.02a$. Unless noted, friction is constant with $\mu\in\{0,1\}$. HS = Hard Sphere}
\begin{tabular}{lccc}
\hline\hline
Case & $F_A/F_0$ & $F_R/F_0$ & friction \\
\hline
Baseline (HS) & $0$ & $0$ & $\mu=0$ or $1$ \\
A0.1 & $0.1$ & $0$ & $\mu=0$ or $1$ \\
A1   & $1$   & $0$ & $\mu=0$ or $1$ \\
A10  & $10$  & $0$ & $\mu=0$ or $1$ \\
R100  & $0$ & $100$  & $\mu=0$ or $1$ \\
R1000 & $0$ & $1000$ & $\mu=0$ or $1$ \\
LD-$\mu$ & $0$ & $0$ & $\mu(F_n)$ (Eq.~(1)) \\
\hline\hline
\end{tabular}
\label{tab:interaction_cases}
\end{table}

\noindent\paragraph*{\textbf{Microscopic interactions and shear-thinning rheology.}} Fig.~\ref{fig:fig1}(b) and (c) show the relative viscosity $\eta_r=\eta/\eta_0$, where $\eta$ is the suspension viscosity, as a function of the dimensionless shear rate $\dot{\gamma}/\dot{\gamma}_0$ for frictionless ($\mu=0$) and frictional ($\mu=1$) particles, respectively. In all cases, the suspensions exhibit pronounced shear thinning over several decades in $\dot{\gamma}$. The magnitude and shear-rate dependence of the $\eta_r$, however, vary strongly with the microscopic interaction mechanism. Strongly attractive systems display large $\eta_r$ at low shear rates followed by substantial thinning, whereas purely repulsive systems exhibit a comparatively weaker shear-rate dependence. Suspensions with load-dependent friction show intermediate behaviour, reflecting the gradual weakening of frictional resistance with increasing shear. Frictional systems have higher viscosities as expected, and the repulsive suspensions undergo continuous shear thickening beyond a critical shear rate, as in this case, the same as the critical load model responsible for shear thickening \cite{mari2014shear}. Interaction-specific differences are further reflected in the normal stress differences, shown in the Supplemental Material (Fig.~S2). 

Fig.~\ref{fig:fig1} establishes that qualitatively different microscopic interaction mechanisms can produce similar shear-thinning trends while generating markedly different stress magnitudes and anisotropies. These observations highlight the central role of microstructure and contact mechanics in setting the rheological response. In the following sections, we examine whether these pronounced differences in stress and anisotropy are mirrored at the level of particle-scale dynamics, or whether the kinematics of particle motion exhibit a more universal behavior across interaction models.

\begin{figure}[t!]
\includegraphics[width=0.7\textwidth]{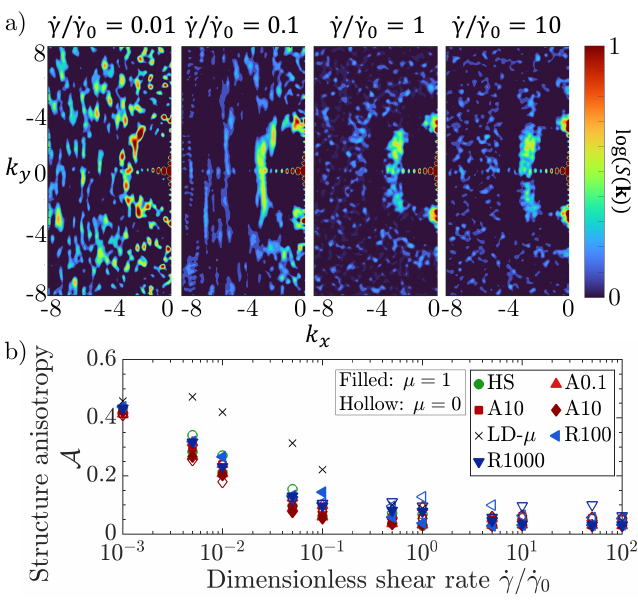}
\vspace{-3mm}
\caption{\label{fig:fig2} Shear-induced structural isotropy in dense suspensions.
(a) Two-dimensional projections of the structure factor $S(\mathbf{k})$ in the flow--gradient plane for increasing dimensionless shear rate $\dot{\gamma}/\dot{\gamma}_0$ for strongly attractive suspension A10. Strong anisotropy is observed at low shear rates and progressively diminishes as shear rate increases.
(b) Scalar structural anisotropy parameter $A$ as a function of $\dot{\gamma}/\dot{\gamma}_0$ for different interaction models and for frictionless ($\mu=0$, open symbols) and frictional ($\mu=1$, filled symbols) particles. Structural anisotropy decreases monotonically with shear rate across all cases, indicating shear-driven loss of microstructural order.}
\vspace{-4mm}
\end{figure}

\noindent\paragraph*{\textbf{Universal shear-induced structural anisotropy results in shear-thinning.}}
Fig.~\ref{fig:fig2} quantifies the evolution of microstructural anisotropy across interaction models and shear rates. Fig.~\ref{fig:fig2}(a) shows two-dimensional projections of the static structure factor $S(\mathbf{k})$ in the flow--gradient plane for representative shear rates for the A10 case. Similar trends are observed for all other interactions (see SI). At low shear rates, the structure factor exhibits pronounced anisotropy, with enhanced scattering along preferred directions reflecting shear-induced isotropy. As the shear rate increases, these anisotropic features progressively weaken, and the structure factor becomes increasingly isotropic, indicating a loss of long-range structural organisation. Thus, Fig.~\ref{fig:fig2}(a) quantifies, in a unified manner, the long-standing qualitative picture that shear thinning in dense suspensions is accompanied by a progressive loss of microstructural anisotropy through flow alignment and structural homogenization.

To make this observation quantitative, Fig.~\ref{fig:fig2}(b) reports a scalar anisotropy measure $\mathcal{A}$, extracted from the angular variation of $S(\bm{k})$. $\mathcal{A}(k) =
\frac{\mathrm{Var}_{\theta}\!\left[S(k,\theta)\right]}
{\langle S(k,\theta)\rangle_{\theta}^2}$, where $\langle\cdot\rangle_{\theta}$ denotes an average over $\theta$. Across all fourteen interactions, $\mathcal{A}$ decreases monotonically with increasing shear rate, demonstrating that shear systematically erodes structural anisotropy irrespective of the underlying microscopic force law. The functional dependence on shear rate strikingly collapses on a single curve irrespective of attraction or repulsion for frictionless ($\mu=0$), frictional ($\mu=1$), and load-dependent friction systems.

\begin{figure}[t]
\includegraphics[width=0.9\textwidth]{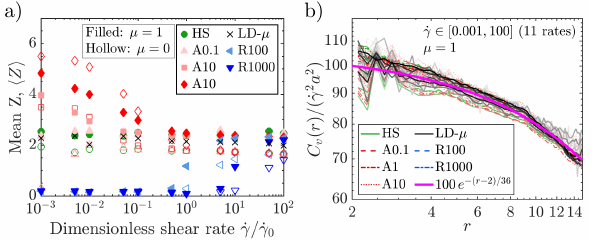}
\vspace{-3mm}
\caption{\label{fig:fig3}Coordination number and velocity correlations under shear.
(a) Mean coordination number $\langle Z\rangle$ as a function of dimensionless shear rate for different interaction models. Coordination decreases with shear rate in an interaction-dependent manner.
(b) Normalized spatial velocity correlation $C_v(r)$ for frictional systems ($\mu=1$), for various shear rates denoted by increasing shade of the line. Despite large variations in microscopic coordination number, velocity correlations exhibit a similar spatial decay across interaction models, indicating emergent kinematic organization at mesoscale.}
\vspace{-4mm}
\end{figure}

\begin{figure*}[t]
\includegraphics[width=\textwidth]{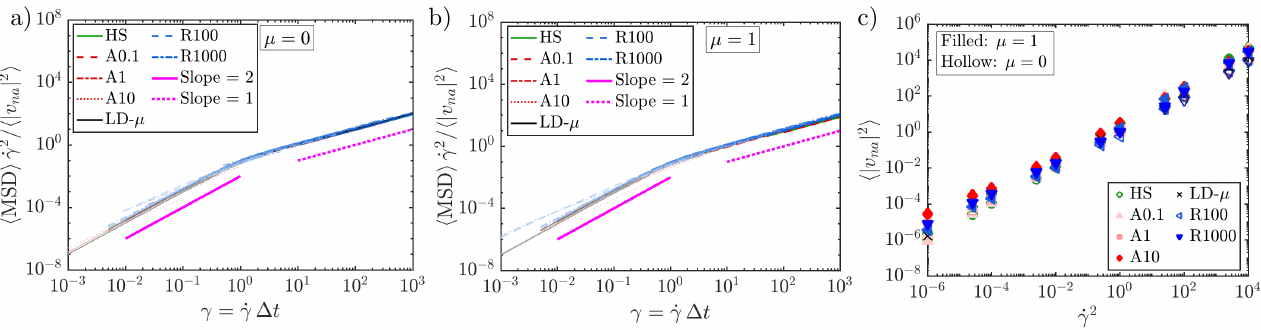}
\vspace{-3mm}
\caption{\label{fig:fig4} Strain--transport superposition of particle motion across interaction models and shear rates.
(a,b) Mean-squared displacement in the gradient--vorticity ($yz$) plane, normalized by the non-affine velocity variance, plotted against accumulated strain $\gamma=\dot{\gamma}\Delta t$ for frictionless ($\mu=0$) and frictional ($\mu=1$) systems, respectively. 
Fourteen distinct interaction models are shown, spanning hard-sphere, attractive (A0.1, A1, A10), repulsive (R100, R1000), and load-dependent friction (LD-$\mu$) suspensions, over eleven shear rates in the range $\dot{\gamma}=10^{-3}$--$10^{3}$. 
Panels (a) and (b) contain data that overlap but are displayed separately to avoid visual overcrowding.
Different shear rates are indicated by progressively darker shading; however, the curves collapse so closely that individual shades are largely indistinguishable, highlighting the robustness of the superposition.
Thick solid and dotted lines indicate reference slopes corresponding to ballistic ($\sim\gamma^{2}$) and diffusive ($\sim\gamma$) regimes, respectively.
(c) non-affine velocity variance $\langle|\bm{v}_{\mathrm{na}}|^2\rangle$ plotted against $\dot{\gamma}^2$ for all interactions, demonstrating the scaling that underpins the strain--transport superposition. 
Filled (hollow) symbols correspond to $\mu=1$ ($\mu=0$), and crosses denote load-dependent friction.
}
\vspace{-4mm}
\end{figure*}

This collapse of $\mathcal{A}(\dot{\gamma})$ is non-trivial: it occurs despite large differences in coordination number, stress transmission, and viscosity (Fig.~\ref{fig:fig1},~\ref{fig:fig3}), and despite the fact that these interaction models generate shear thinning through distinct microscopic mechanisms. Fig.~\ref{fig:fig2} therefore provides the first systematic evidence that the reduction of structural anisotropy with shear rate is a robust, interaction-independent feature of shear-thinning dense suspensions, placing the traditional microstructural explanation of shear thinning on a quantitative footing.

Fig.~\ref{fig:fig2} also highlights a limitation of a purely structural interpretation. Although microstructural anisotropy collapses onto a single curve when plotted against shear rate, the corresponding rheological responses do not. As we show next, this decoupling becomes even more pronounced at the level of particle-scale dynamics: despite substantial and universal changes in microstructure, the kinematics of particle motion obey an even stronger form of universality, governed by strain accumulation rather than interaction-specific structural relaxation.

\paragraph*{\textbf{Velocity correlations and coordination under shear.}} Fig.~\ref{fig:fig3} examines the relationship between microstructural connectivity and particle-scale kinematics under steady shear. Fig.~\ref{fig:fig3}(a) shows the average coordination number $\langle Z\rangle$ as a function of dimensionless shear rate for all interaction models. $\langle Z\rangle$ decreases with increasing shear rate, reflecting progressive contact loss and the erosion of the load-bearing network, except for the repulsive suspensions. The magnitude and rate of this decrease depend strongly on the microscopic interaction mechanism: attractive interactions sustain higher coordination at low shear rates, whereas repulsive interactions exhibit significantly reduced coordination, as expected. Load-dependent friction displays intermediate behavior, consistent with its gradual weakening of frictional resistance under increasing shear.

Fig.~\ref{fig:fig3}(b) shows the spatial velocity correlation function $C_v(r)$ for all interaction models at $\mu=1$, averaged over the full range of shear rates studied. Despite the pronounced differences in coordination number and contact mechanics revealed in Fig.~\ref{fig:fig3}(a), the spatial decay of velocity correlations is remarkably similar across interaction types. When appropriately rescaled, $C_v(r)$ collapses onto a common curve over a broad range of separations, indicating that the spatial organization of particle velocities is largely insensitive to the details of the interaction potential. The dashed line indicates an empirical fit capturing the dominant decay of correlations with distance.

Fig.~\ref{fig:fig1}-\ref{fig:fig3} highlight a striking contrast: while microscopic coordination number and macroscopic rheology vary strongly with shear rate and interaction model, the spatial mesoscale structure of velocity correlations remains similar across interactions and shear rates. This observation suggests that microscopic changes primarily regulate force transmission rather than the kinematics of particle motion, thus hinting towards an emergent kinematic universality. In the next section, we demonstrate that this insensitivity to micromechanics at the mesoscale extends to particle transport, showing that non-affine velocity fluctuations set a universal scale for mean-square displacements across all interaction models.


\paragraph*{\textbf{Strain--transport superposition.}}
Fig.~\ref{fig:fig4} reveals a central organizing principle for shear-driven dynamics in shear-thinning dense non-Brownian suspensions: \emph{strain--transport superposition}. Fig.~\ref{fig:fig4}(a) shows the mean-squared displacement (MSD) transverse to the flow (in the $yz$ plane), rescaled by the non-affine velocity variance and plotted versus accumulated strain $\gamma=\dot{\gamma}\Delta t$, for fourteen interaction models and eleven shear rates spanning six decades. Focusing on transverse displacements removes trivial affine advection and isolates irreversible, rearrangement-driven transport. Despite large variations in microscopic interactions, coordination number, and rheological response, all data collapse onto a single master curve.

At small accumulated strain, the MSD grows ballistically, $\mathrm{MSD}\sim\gamma^{2}$, reflecting persistence of non-affine particle velocities over short deformation intervals. At larger strain, successive shear-driven encounters decorrelate these velocities, producing diffusive transport with $\mathrm{MSD}\sim\gamma$. The crossover occurs at $\gamma \approx \mathcal{O}(1)$ for all interaction models, demonstrating that accumulated strain—rather than stress, coordination number, or microstructural anisotropy—sets the decorrelation scale for transverse particle motion.

This crossover is directly quantified by the strain-dependent autocorrelation of the non-affine velocity. The normalized three-dimensional non-affine velocity autocorrelation,
$C_{v_{\mathrm{na}}}(\Delta\gamma)
=
\frac{
\left\langle
\bm{v}_{\mathrm{na}}(t)\cdot \bm{v}_{\mathrm{na}}(t+\Delta t)
\right\rangle
}{
\left\langle
|\bm{v}_{\mathrm{na}}|^2
\right\rangle
},
$
decays sharply to 0 at $\Delta\gamma\sim\mathcal{O}(1)$ for all shear rates and interaction mechanisms (see SI). This rapid loss of velocity memory provides a microscopic explanation for the universal ballistic--to--diffusive crossover in the MSD: persistent non-affine motion dominates at small strain, while decorrelation beyond one unit of accumulated strain yields diffusive dynamics. The collapse of the autocorrelation curves when plotted against strain confirms that decorrelation is controlled by deformation rather than by time, stress, or interaction-specific relaxation scales.

This behavior follows from simple scaling arguments. In overdamped suspensions, particle velocities are set by instantaneous force balance, and the only imposed kinematic scale is the shear rate. Hydrodynamic forcing scales as $F_H\sim\eta_0\dot{\gamma}a^2$, implying $|\bm v_{\mathrm{na}}|\sim\dot{\gamma}a$ in the absence of additional intrinsic time scales. Fig.~\ref{fig:fig4}(b) confirms this expectation, showing $\langle|\bm v_{\mathrm{na}}|^2\rangle\propto\dot{\gamma}^2$ across all interaction models. Together, Fig.~\ref{fig:fig4}(a,b) imply the universal form
\[
\mathrm{MSD}_{yz}(\Delta t)
=
\frac{\langle|\bm v_{\mathrm{na}}|^2\rangle}{\dot{\gamma}^2}\,
F(\dot{\gamma}\Delta t),
\]
with a scaling function $F$ independent of microscopic interaction details. We find $F=0.1\,\gamma^2$ for $\gamma\lesssim\mathcal{O}(1)$ and $F=0.1\,\gamma$ for $\gamma\gtrsim\mathcal{O}(1)$, consistent with classical measurements of shear-induced diffusion where dimensionless diffusivities are typically $\mathcal{O}(10^{-1})$ in dense non-Brownian systems~\cite{LeightonAcrivos1987JFM,BradyMorris1997JFM}. The persistence of the same prefactor across all interactions confirms that it is a kinematic property of dense shear-driven motion rather than a signature of interaction-specific force transmission.

Strain--transport superposition is thus analogous to time--temperature superposition in polymers~\cite{more2023rod, more2024elasto}: the shear rate reparametrizes progression along a common dynamical pathway without altering the underlying transport mechanism. Supplemental analyses (SI) show that while the dimensionless amplitude $\langle|\bm v_{\mathrm{na}}|^2\rangle/\dot{\gamma}^2$ exhibits modest interaction-dependent variations, particularly at low $\dot{\gamma}$, these variations do not correlate uniquely with viscosity, unlike behavior near jamming in shear-thickening suspensions~\cite{singh2023scaling}. Instead, interaction details primarily govern how strain-driven motion is converted into stress through dissipation and force transmission.

Taken together, Fig.~\ref{fig:fig4} reconciles the rheological and structural trends in Figs.~\ref{fig:fig1}--\ref{fig:fig3}. Although microstructure, coordination number, and stress anisotropy vary strongly with interaction model and shear rate, the transverse kinematics of particle motion obey strain--transport superposition. Shear thinning in dense non-Brownian suspensions therefore reflects a renormalization of stress transmission at fixed, strain-controlled kinematics, with non-affine velocity fluctuations emerging as the fundamental dynamical scale governing shear-driven transport.

\paragraph*{\textbf{Conclusions.}} We have demonstrated a fundamental decoupling between particle-scale kinematics and macroscopic stress in dense non-Brownian suspensions under shear. Despite large variations in interaction mechanisms, coordination number, and stress anisotropy, the magnitude of non-affine particle velocities is universally set by the imposed shear rate. When transport is expressed in strain units and rescaled by the non-affine velocity variance, particle motion collapses onto a single universal form with a robust ballistic-to-diffusive crossover. These results identify non-affine velocity fluctuations as the emergent kinematic scale governing shear-driven transport, independent of the microscopic origin of shear thinning, and suggest that shear controls particle motion while microstructure determines the force required to sustain it. This separation provides a unifying framework for transport, mixing, and irreversibility in dense particulate flows.

\begin{acknowledgments}
\textit{\textbf{Acknowledgments}.} The author thanks Prof. Abhi Singh for insightful discussions, Monash Massive supercomputing resources and support from the Engineering Talent Research Accelerator fellowship by the Faculty of Engineering at Monash University. 
\end{acknowledgments}

\vspace{-4mm}
\appendix

\section{Appendixes}

\paragraph{Methodology.} We perform three-dimensional particle-scale simulations of dense non-Brownian suspensions in the Stokes-flow regime (Reynold number $\mathrm{Re}\!\ll\!1$), where particle motion is governed by instantaneous force balance between hydrodynamic interactions and non-hydrodynamic particle forces. The system contains $N=500$ neutrally buoyant, bidisperse spheres (radius ratio $1.4$ with equal volume fractions) at total volume fraction $\phi=0.5$ in a periodic cubic domain of size $L=18.31a$ (with $a$ the radius of the smaller species). A steady simple shear flow
$\bm{v}_{\mathrm{aff}}=\dot{\gamma}y\,\hat{\bm{x}}$
is imposed using Lees--Edwards boundary conditions ($x$: flow, $y$: gradient, $z$: vorticity). Brownian motion is negligible as Péclet number $\mathrm{Pe}\gg 1$. Governing equations, interaction laws, numerical algorithm, and validation are provided in the Supplemental Materials (SI) and have been extensively validated in our prior works~\cite{more2020effect, more2020roughness, more2020constitutive, more2021unifying}.

\paragraph{Microstructural metrics.}
We characterize microstructure and dynamics by computing the static structure factor $S(\bm{k})$ (the Fourier transform of the pair distribution function $g(\bm{r})$), the mean coordination number $\langle Z\rangle$, spatial velocity correlations $C_v(r)$, non-affine velocity fluctuations $\bm{v}{\mathrm{na}}$, and mean-squared displacements (MSDs). non-affine velocities are defined relative to the imposed affine shear as $\bm{v}_{\mathrm{na}}=\bm{U}-\dot{\gamma}y\hat{\bm{x}}$, where $\bm{U}$ is the particle velocity, and their magnitude is quantified by the variance $\langle|\bm{v}_{\mathrm{na}}|^2\rangle$. Shear-induced structural anisotropy is analyzed from two-dimensional projections of $S(\bm{k})$ in the $xy$, $xz$, and $yz$ planes and condensed into a scalar \emph{structure-factor anisotropy metric} $\mathcal{A}$, defined via the angular variance of $S$ over a prescribed $k$-band (SI). While anisotropic scattering under shear is widely used qualitatively (and sometimes via simple intensity ratios) in rheo-scattering and confocal studies~\cite{Cheng2011Science}, we use a single, plane-resolved scalar measure to enable systematic comparison across distinct microscopic thinning mechanisms and to place microstructural anisotropy on the same footing as kinematic observables. MSDs are evaluated in the gradient--vorticity ($yz$) plane as functions of accumulated strain to isolate transverse (irreversible) transport. All quantities are averaged in steady state over the final $10^2$ strain units; detailed expressions are given in the Supplemental Material (SI).


\nocite{*}
\bibliographystyle{apsrev4-1}
\bibliography{apssamp}

\end{document}